\documentclass[prl,preprint,floatfix,superscriptaddress,nofootinbib]{revtex4-1} 
\usepackage{graphicx}
\usepackage{natbib}
\usepackage{mathrsfs} 

\newlength{\onecolfig}
\newlength{\twocolfig}
\newcommand{\unit}[1]{\,\mbox{#1}}

\newcommand{\Hz}{\unit{Hz}}

\newcommand{\GHz}{\unit{GHz}}

\newcommand{\T}{\unit{T}}

\newcommand{\um}{\unit{$\mu$m}}

\newcommand{\K}{\unit{K}}

\newcommand{\Ctsps}{\unit{Cts/s}}
\newcommand{\ms}{\unit{ms}}

\newcommand{\scL}{\ensuremath{\mathrm{\mathscr{L}}}}

\begin{document}

\title{A charge-driven feedback loop in the resonance fluorescence of a single quantum dot}

\author{B.~Merkel}
\affiliation{University of Duisburg-Essen, Faculty of Physics and CENIDE, Lotharstr. 1, 47057 Duisburg, Germany}
\author{A.~Kurzmann}
\affiliation{University of Duisburg-Essen, Faculty of Physics and CENIDE, Lotharstr. 1, 47057 Duisburg, Germany}

\author{J.-H.~Schulze}
\affiliation{Technische Universit\"{a}t Berlin, Institut f\"{u}r Festk\"{o}rperphysik, Hardenbergstrasse 36, 10623 Berlin, Germany}
\author{A.~Strittmatter}
\affiliation{Technische Universit\"{a}t Berlin, Institut f\"{u}r Festk\"{o}rperphysik, Hardenbergstrasse 36, 10623 Berlin, Germany}

\author{M.~Geller}
\affiliation{University of Duisburg-Essen, Faculty of Physics and CENIDE, Lotharstr. 1, 47057 Duisburg, Germany}
\author{A.~Lorke}
\affiliation{University of Duisburg-Essen, Faculty of Physics and CENIDE, Lotharstr. 1, 47057 Duisburg, Germany}

\email{martin.geller@uni-due.de}

\date{\today}

\begin{abstract}
Semiconductor quantum dots can emit antibunched, single photons on demand with narrow linewidths \cite{Michler.2000, He.2013, Kuhlmann.2015}. However, the observed linewidths are broader than lifetime measurements predict, due to spin and charge noise in the environment \cite{Kuhlmann.2013,Matthiesen.2014}. This noise randomly shifts the transition energy and destroys coherence \cite{Muller.2007,Flagg.2009} and indistinguishability \cite{Santori.2002,Matthiesen.2013} of the emitted photons. Fortunately, the fluctuations can be reduced by a stabilization using a suitable feedback loop \cite{Latta.2009,Prechtel.2013,Hansom.2014}. In this work we demonstrate a fast feedback loop that manifests itself in a strong hysteresis and bistability of the exciton resonance fluorescence signal. Field ionization of photogenerated quantum dot excitons leads to the formation of a charged interface layer that drags the emission line along over a frequency range of more than $30\GHz$. This internal charge-driven feedback loop could be used to reduce the spectral diffusion and stabilize the emission frequency within milliseconds, presently only limited by the sample structure, but already faster than nuclear spin feedback. 
\end{abstract}

\maketitle

A vision in quantum information technology is a quantum network \cite{Bennett.2000,Kimble.2008} where the quantum bits (qubits) are stored in nodes and connected via quantum channels. Semiconductor quantum dots (QDs) are one possible candidate for such a quantum node \cite{Ladd.2010}. They can host, for instance, a spin qubit \cite{Vamivakas.2009}, and the quantum information can be transferred between different nodes via single photons \cite{Ylmaz.2010}. However, these photons suffer from frequency noise, induced by residual charges and spins in their environment \cite{Kuhlmann.2013, Matthiesen.2014}, which reduces their coherence and indistinguishability \cite{Santori.2002,Matthiesen.2013}. 

A possibility to reduce this spectral jitter is a feedback loop that counteracts the external fluctuations. One option is an {\it external} feedback that measures the emission frequency and then corrects it by setting a gate voltage, using the Stark effect\cite{Prechtel.2013,Hansom.2014}. For example, the resonance fluorescence (RF) of a self-assembled QD produces single photons while simultaneously the emission energy is determined from the differential transmission \cite{Prechtel.2013}. Such a feedback loop effectively stabilizes the QD photon emission up to a frequency of about $100\Hz$. Alternatively, the feedback can be based on {\it internal} mechanisms, using the coupling of the electron-hole pairs to the environment. In magnetic fields, the interaction of the excitons with the nuclear spins in the dot leads to a dragging-like hysteresis of the resonance fluorescence. The hysteresis is caused by dynamic nuclear spin polarization \cite{Hogele.2012, Chekhovich.2013} and can also be used to stabilize the emission line \cite{Latta.2009,Yang.2013}. Such an internal feedback loop needs no further optical or electrical detection scheme. However, the bandwidth of this internal nuclear spin feedback seems to be restricted to about $10\Hz$.  

\begin{figure*}
\includegraphics[width=0.9\twocolfig]{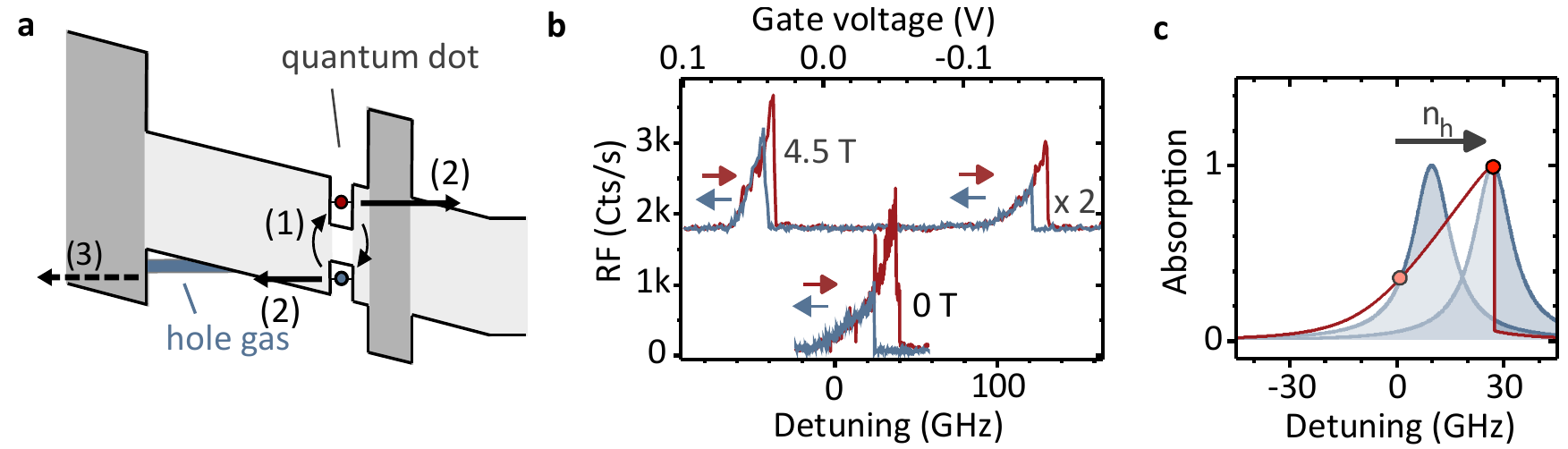}
\caption{{\bf The formation of a hole gas shifts the exciton resonance.} 
{\bf a}, Band structure of the active region of the device (see also methods section and supplementary information). AlGaAs and GaAs layers are shades in dark and grey, respectively. Arrows indicate the processes involved in the hole gas formation: (1) Exciton generation in the QD, (2) tunnelling of electron and hole, (3) and leakage of the hole gas. {\bf b}, Measured fluorescence for increasing (red) and decreasing (blue) gate voltage at $0\T$ and $4.5\T$ (shifted vertically by $1800 \Ctsps$ for clarity). The detuning is given with respect to the exciton resonance in the absence of a hole gas (deduced from a fit of the line shape to our rate equation model). 
{\bf c}, Schematic illustration of two absorption lines (in blue) for two fixed hole gas populations $n_h$. The solid red line depicts schematically the measured fluorescence curve when the absorption line is dragged along by the feedback of the hole gas. 
}
\label{fig-1}
\end{figure*}

We use here a micro-patterned device that consists of a Schottky diode with a layer of self-assembled InAs/GaAs QDs (see the schematic band structure in Fig.~\ref{fig-1}a and methods below) to demonstrate an inherently fast internal and charge-driven feedback loop in the resonance fluorescence of a single dot. The bottom trace in Fig.~\ref{fig-1}b shows the RF signal when the exciton transition is shifted upwards or downwards across the laser frequency (blue and red lines, respectively) by sweeping the voltage that is applied to the device. Compared to the commonly found Lorentzian line shape of excitons in single InAs quantum dots \cite{Kuhlmann.2015}, the resonances observed here show several distinctly different characteristics. They are much broader and exhibit a strongly asymmetric shape with a gradual increase on the low energy side and an abrupt decrease on the high energy side. More importantly, the traces show a pronounced hysteresis over a range of $15\GHz$ not only in the absence of magnetic field ($B= 0$) but also for both Zeeman-split exciton transitions at $B=4.5\T$. 

The width and the asymmetry of the resonance as well as the hysteresis indicate that the resonance is ``dragged along'' as the excitation frequency is shifted. This can be explained by a three-step process that leads to the buildup of transient charge at the GaAs/(AlGa)As interface, as indicated in Fig.~\ref{fig-1}a \cite{Hosoda.1996, Luttjohann.2005, Kroner.2008, Hauck.2014}: (1) Upon resonant excitation, bound electron hole pairs are generated in the dot. (2) These excitons can be field ionized by tunnelling \cite{Seidl.2005}. The electron escapes to the back contact, while the hole becomes trapped in the triangular well at the GaAs-(AlGa)As interface. (3) The stored, positive charge in that well will slowly drain, e.\,g. by tunnelling through the (AlGa)As barrier. Process (3) depends on the total number of holes in the well, so that for a given pumping efficiency (process (2)), a specific amount of transient positive charge will be stored at the interface. This charge affects the electric field across the dot, which will slightly detune the resonance with respect to the laser frequency.

The resulting feedback loop (pumping efficiency $\rightarrow$ amount of trapped charge $\rightarrow$ resonance position $\rightarrow$ pumping efficiency) can explain the resonance width and shape as well as the bistability: When the pumping frequency is on the low energy side of the exciton resonance line (red point at zero detuning in Fig.~\ref{fig-1}c), the feedback will be negative, so that a stable configuration is achieved: A small increase in the number $n_h$ of stored holes will blue-shift the resonance to more positive detuning, as indicated by the arrow. This will reduce the overlap of the exciton line with the laser, reduce the pumping efficiency, and decrease $n_h$ again. Correspondingly, there will be a positive feedback on the high energy side of the resonance, which explains the abrupt drop of the fluorescence on this side. When the laser coincides with the maximum of the exciton line (red point at 30 GHz detuning in Fig.~\ref{fig-1}c), a small decrease in $n_h$ will red-shift the resonance. This will further reduce the pumping efficiency until the resonance has almost completely shifted out of resonance. 

\begin{figure}
\includegraphics[width=0.9\onecolfig]{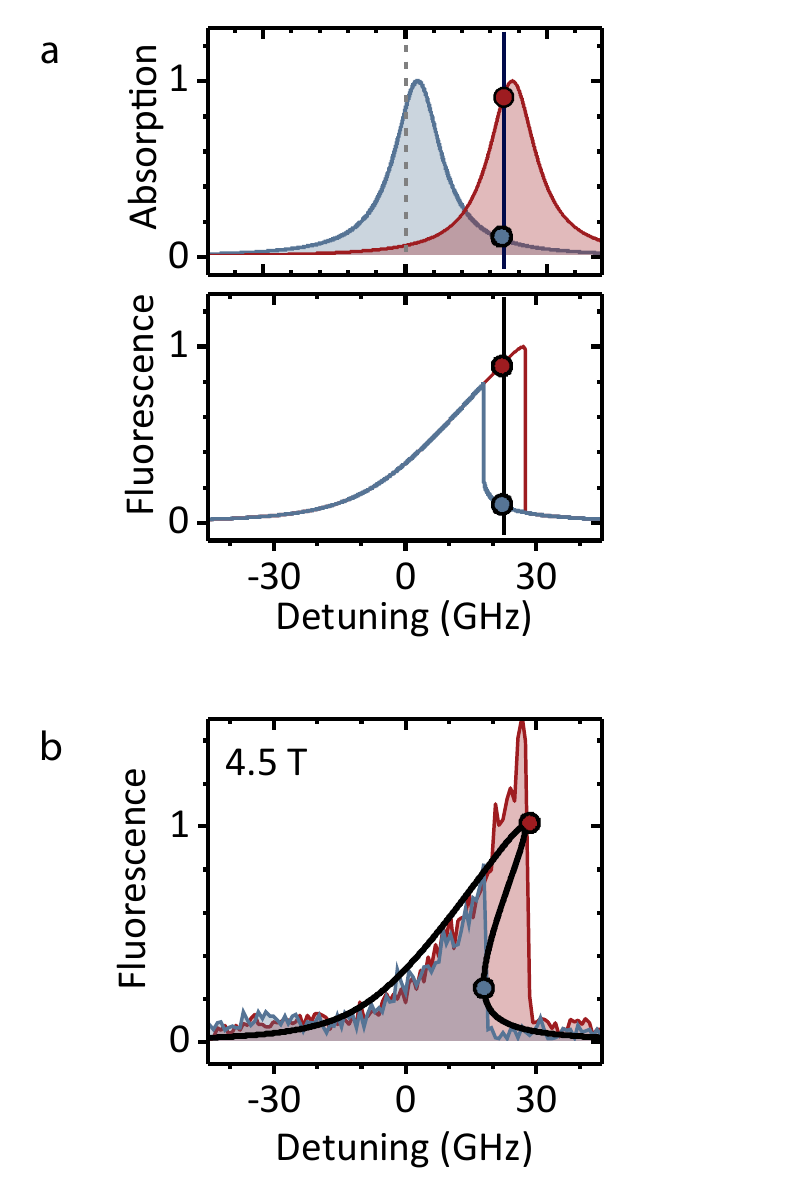}
\caption{{\bf Explanation of the optical bistability and hysteresis curve.} 
{\bf a}, The upper panel shows schematically the resonance line for two different hole gas populations. The blue absorption curve exists for a small hole gas population. The red curve is an example for a higher population, e.~g.~shifting the resonance to more positive detuning. As a consequence, an excitation laser with fixed frequency (black vertical line) can drive the transition on the high energy side (blue solid dot) and on the low energy side (red solid dot). The resonance florescence signal will be different in this region of bistability and a strong hysteresis is observed  (see lower panel for calculated fluorescence curves). 
{\bf b}, The measured resonance fluorescence for increasing (red line) and decreasing detuning (blue line) is compared with the stationary solutions of a rate equation model (black line, see supplementary information for details on the model).
}
\label{fig-2}
\end{figure}

Just below the maximum, there will be two possible stable configurations, as depicted in Fig.~\ref{fig-2}a. For a more quantitative description of the observed line shape and hysteresis, we model the system by a rate equation approach. The number $n_h$ of trapped holes is given by the interplay between the processes (2) and (3), described by the rates $\gamma_{pump}$ and $\gamma_{leak}$, respectively. $\gamma_{pump}$ is determined by an ionization probability $P_{ionize}$ and the overlap between the laser line and the Lorentzian line shape $\scL(n_h)$ of the exciton absorption line, with a position that depends on $n_h$. The leakage rate is assumed to be proportional to $n_h$. The resulting differential equation
\begin{equation}
\frac{d n_h}{dt}=\gamma_{pump}-\gamma_{leak}= \gamma_{abs}\;\scL(n_h)\;P_{ionize} -\gamma_{esc} n_h
\end{equation}
is solved numerically, taking the proportionality factors $\gamma_{abs}$, $P_{ionize}$ and $\gamma_{esc}$ as experimental parameters. For more details on the modelling, see the supplementary material.  The stationary solutions of Eq. 1 are shown as a black line in Fig.~2b. Our model can well account for all characteristics of the observed resonances. The hysteresis regime of the fluorescence curves is given by the fold-over region (between $20$ and $30\GHz$ in  Fig.~\ref{fig-2}b) in which three stationary solutions for a fixed detuning exist. Only two solutions are stable, corresponding to the observed fluorescence states, and switching occurs when the edge of the fold-over region is reached (solid dots). As already suggested above, the broad, triangular shape of the resonance results from the dragging-along of the resonance.

So far, the observed phenomena share some characteristics with the recently observed nuclear-spin-induced dragging and the resulting hysteresis \cite{Latta.2009,Hogele.2012}. In the present system, however, the feedback is based on photoionization in electric fields rather than magnetic interaction with nuclear spins, which makes the response orders of magnitude more rapid and allows for fast manipulation by externally applied voltage or light pulses. This will be demonstrated in the following.

\begin{figure*}
\includegraphics[width=0.85\twocolfig]{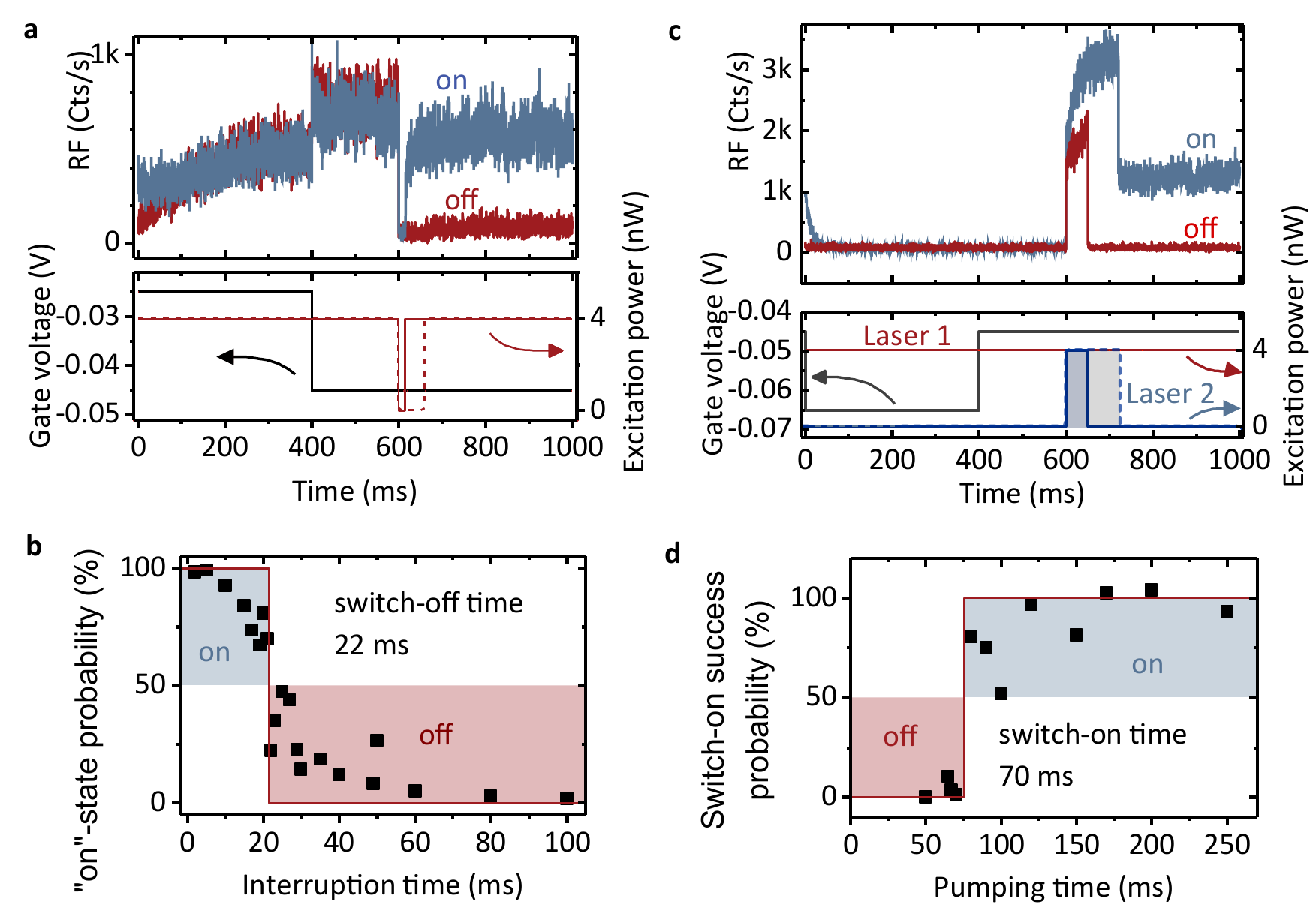}
\caption{{\bf Time-resolved optical switching between the ''on''- and ''off''-state in the bistability region.}  {\bf a}, The lower panel shows the switch-off sequence for the gate voltage (black line) and laser excitation power (red line). At $t= 400\ms$, the gate voltage is switched to a position in the hysteresis region. The QD is now in resonance with the excitation laser (region between $20-30$ GHz in Fig.~\ref{fig-2}b) and a strong RF signal is visible in the upper panel. At $t= 600$ ms, the laser is interrupted for a variable duration and the QD resonance will be red-shifted as the stabilization by the hole gas population is switched off. Depending on the interruption time, the dot will be in the ''on''- or ''off''-state after switching the laser on again (blue and red RF traces in the upper panel, respectively).  {~\bf b}, Probability of recovering the QD fluorescence as a function of the interruption time. {\bf c}, (lower panel) Switch-on sequence for gate voltage (black line) and laser power. Laser 1 (red line) is set to the hysteresis region, while laser 2 (blue line) is in resonance with the unshifted transition. Switching on laser 2 at $t=600$~ms blue-shifts the transition until laser 1 is within the RF line width. Depending on the pumping time of laser 2, the system is in the ''on''- or in the ''off''-state (blue and red line in the upper panel, respectively) after switching off the second laser. {\bf d}, Success probability as a function of the pumping time with laser 2.}
\label{fig-3}
\end{figure*}

Figure~\ref{fig-3}a shows how the resonance fluorescence in the hysteresis region is affected by an interruption of the laser excitation. The quantum dot is first initialized in the high-fluorescence ``on''-state by laser irradiation and application of a suitable gate voltage. Then, the gate voltage is switched to the hysteresis region (at time $t= 400\ms$ in Fig.~\ref{fig-3}a). At $600 \ms$, the laser excitation is interrupted for a short, variable time span $\Delta t$. Subsequently, it is recorded whether the system is in the ``on''-state (blue curve) or the ``off''-state (red curve). In Fig.~\ref{fig-3}b, the ``on''-state probabilities are plotted as a function of the interruption time $\Delta t$. We observe a latency time (time required for the ``on''-state probability to drop below 50\,\%) of roughly $22\ms$.  

Similarly, the fluorescence can be switched on by optical means, using a second laser. Laser~1 is set to the hysteresis region and the QD is initialized in the ``off''-state (Fig.~\ref{fig-3}c). Then, for a short pumping time $\Delta t$, laser~2 is turned on, which is resonant with the initial exciton frequency. This populates the hole gas and shifts the transition to higher frequencies, closer to laser~1 in the hysteresis region. Depending on the length of the additional laser pulse, the system will afterwards be in the ``off'' or ``on''-state with respect to laser~1 (red and blue traces in Fig.~\ref{fig-3}c, respectively). Again, a latency period can be observed, which is $70\ms$ for the present experimental conditions. 

The time dependence of the switch-off process is given by the leakage rate of the hole gas (process (3) in Fig.~1a). Note, however, that for very short interruption times $\Delta t$, the negative feedback mechanism will ensure a fast recovery of the RF signal. Only for sufficiently long $\Delta t$, the resonance position will have shifted far enough that a recovery is no longer possible. This explains the observed latency time. Correspondingly, the observed latency period of the switch-on process reflects the pumping rate of the hole gas by exciton generation and field ionization. It is worth noting that the particular characteristic times strongly depend on the excitation power and the position of the laser frequency within the bistability region.

The operation principle of our feedback loop, which is based on simple charge transport and therefore offers great tunability, opens up a broad variety of possibilities. Similar to the proposal based on the nuclear dragging effect, the charge driven feedback loop may be used for stabilization of the QD resonance position. Also, the bistability may be useful for optical memory applications, such as switching the fluorescence signal from  the ``off'' to the ``on''-state when a short optical signal is received and holding that state for later read-out. 


\section*{Methods}
\subsection*{Sample growth and processing}
The investigated quantum dots were grown using the now well-established Stranski-Krastanov growth mode and consist of InAs, embedded in a GaAs/(AlGa)As field-effect-transistor-like heterostructure  \cite{Warburton.2000}. The electrically and optically active part of the sample is sketched in Fig.~\ref{fig-1}a. It consists (along the growth direction from right to left) of a highly doped, metallic GaAs back contact, a GaAs-(AlGa)As tunnelling barrier, the quantum dot layer, a GaAs spacer layer, and an (AlGa)As blocking layer. In a pre-growth etching process the sample was patterned into cylindrical mesas of $\sim18.6\um$ diameter \cite{Strittmatter.2012,Strittmatter.2012b}. Lithographically defined contacts on top of a single mesa allow for controlling the electric field at the quantum dots by applying a gate voltage. A more detailed description of the layer sequence and the growth and patterning processes can be found in the supplementary information. 

\subsection{Optical measurements}
The optical response of a single quantum dot was probed by resonance fluorescence spectroscopy (see Refs.~\cite{Flagg.2009,Kuhlmann.2013}). Resonance between the fixed laser frequency and the excitonic transition in the dot was achieved by applying a voltage between a gate electrode on top of the heterostructure and the back contact. The resulting change in electric field across the quantum dot leads to a linear Stark shift of the exciton transition \cite{Warburton.2002}. This shift can be calibrated using different laser frequencies and is given in GHz. All experiments were carried out with the sample placed in a helium bath cryostat at $4.2\K$. A confocal dark-field microscope was used to focus the beam of a tunable diode laser onto a single dot and detect its resonance fluorescence with an avalanche photodiode (APD). To increase the collection efficiency, a zirconia solid immersion lens was placed on top of the sample and centered above a mesa. By the use of superconducting magnetic coils inside the bath cryostat parts of the experiments were performed at a magnetic field of $4.5\T$ along the sample growth direction. 
For time-resolved measurements, a function generator was used to apply voltage pulses either directly at the gate contact or to the driver of an acusto-optic modulator, switching a laser beam on and off. The counts of the APD were then binned by a quTAU time-to-digital converter triggered by the pulse generator. 

\section{Authors contribution}
B.~M.~and A.~K.~carried out the experiments. B.~M.~implemented the model and performed the calculations. J.-H.~S. and A.~S. grew the samples. B.~M., A.~K., M.~G. and A.~L. planned the experiments and wrote the manuscript.


\begin{thebibliography}{10}
	\expandafter\ifx\csname url\endcsname\relax
	\def\url#1{\texttt{#1}}\fi
	\expandafter\ifx\csname urlprefix\endcsname\relax\def\urlprefix{URL }\fi
	\providecommand{\bibinfo}[2]{#2}
	\providecommand{\eprint}[2][]{\url{#2}}
	
	\bibitem{Michler.2000}
	\bibinfo{author}{Michler, P.} \emph{et~al.}
	\newblock \bibinfo{title}{A quantum dot single-photon turnstile device}.
	\newblock \emph{\bibinfo{journal}{Science}} \textbf{\bibinfo{volume}{290}},
	\bibinfo{pages}{2282--2285} (\bibinfo{year}{2000}).
	
	\bibitem{He.2013}
	\bibinfo{author}{He, Y.-M.} \emph{et~al.}
	\newblock \bibinfo{title}{On-demand semiconductor single-photon source with
		near-unity indistinguishability}.
	\newblock \emph{\bibinfo{journal}{Nature Nanotech.}}
	\textbf{\bibinfo{volume}{8}}, \bibinfo{pages}{213--217}
	(\bibinfo{year}{2013}).
	
	\bibitem{Kuhlmann.2015}
	\bibinfo{author}{Kuhlmann, A.~V.} \emph{et~al.}
	\newblock \bibinfo{title}{Transform-limited single photons from a single
		quantum dot}.
	\newblock \emph{\bibinfo{journal}{Nature Commun.}}
	\textbf{\bibinfo{volume}{6}}, \bibinfo{pages}{8204} (\bibinfo{year}{2015}).
	
	\bibitem{Kuhlmann.2013}
	\bibinfo{author}{Kuhlmann, A.~V.} \emph{et~al.}
	\newblock \bibinfo{title}{Charge noise and spin noise in a semiconductor
		quantum device}.
	\newblock \emph{\bibinfo{journal}{Nature Phys.}} \textbf{\bibinfo{volume}{9}},
	\bibinfo{pages}{570--575} (\bibinfo{year}{2013}).
	
	\bibitem{Matthiesen.2014}
	\bibinfo{author}{Matthiesen, C.}, \bibinfo{author}{Stanley, M.~J.},
	\bibinfo{author}{Hugues, M.}, \bibinfo{author}{Clarke, E.} \&
	\bibinfo{author}{Atat{\"u}re, M.}
	\newblock \bibinfo{title}{Full counting statistics of quantum dot resonance
		fluorescence}.
	\newblock \emph{\bibinfo{journal}{Sci. Rep.}} \textbf{\bibinfo{volume}{4}},
	\bibinfo{pages}{4911} (\bibinfo{year}{2014}).
	
	\bibitem{Muller.2007}
	\bibinfo{author}{Muller, A.} \emph{et~al.}
	\newblock \bibinfo{title}{Resonance fluorescence from a coherently driven
		semiconductor quantum dot in a cavity}.
	\newblock \emph{\bibinfo{journal}{Phys. Rev. Lett.}}
	\textbf{\bibinfo{volume}{99}}, \bibinfo{pages}{187402}
	(\bibinfo{year}{2007}).
	
	\bibitem{Flagg.2009}
	\bibinfo{author}{Flagg, E.~B.} \emph{et~al.}
	\newblock \bibinfo{title}{Resonantly driven coherent oscillations in a
		solid-state quantum emitter}.
	\newblock \emph{\bibinfo{journal}{Nature Phys.}} \textbf{\bibinfo{volume}{5}},
	\bibinfo{pages}{203--207} (\bibinfo{year}{2009}).
	
	\bibitem{Santori.2002}
	\bibinfo{author}{Santori, C.}, \bibinfo{author}{Fattal, D.},
	\bibinfo{author}{Vuckovi{\'c}, J.}, \bibinfo{author}{Solomon, G.~S.} \&
	\bibinfo{author}{Yamamoto, Y.}
	\newblock \bibinfo{title}{Indistinguishable photons from a single-photon
		device}.
	\newblock \emph{\bibinfo{journal}{Nature}} \textbf{\bibinfo{volume}{419}},
	\bibinfo{pages}{594--597} (\bibinfo{year}{2002}).
	
	\bibitem{Matthiesen.2013}
	\bibinfo{author}{Matthiesen, C.} \emph{et~al.}
	\newblock \bibinfo{title}{Phase-locked indistinguishable photons with
		synthesized waveforms from a solid-state source}.
	\newblock \emph{\bibinfo{journal}{Nature Commun.}}
	\textbf{\bibinfo{volume}{4}}, \bibinfo{pages}{1600} (\bibinfo{year}{2013}).
	
	\bibitem{Latta.2009}
	\bibinfo{author}{Latta, C.} \emph{et~al.}
	\newblock \bibinfo{title}{Confluence of resonant laser excitation and
		bidirectional quantum-dot nuclear-spin polarization}.
	\newblock \emph{\bibinfo{journal}{Nature Phys.}} \textbf{\bibinfo{volume}{5}},
	\bibinfo{pages}{758--763} (\bibinfo{year}{2009}).
	
	\bibitem{Prechtel.2013}
	\bibinfo{author}{Prechtel, J.~H.} \emph{et~al.}
	\newblock \bibinfo{title}{Frequency-stabilized source of single photons from a
		solid-state qubit}.
	\newblock \emph{\bibinfo{journal}{Phys. Rev. X}}
	\textbf{\bibinfo{volume}{041006}}, \bibinfo{pages}{1--7}
	(\bibinfo{year}{2013}).
	
	\bibitem{Hansom.2014}
	\bibinfo{author}{Hansom, J.}, \bibinfo{author}{Schulte, C. H.~H.},
	\bibinfo{author}{Matthiesen, C.}, \bibinfo{author}{Stanley, M.~J.} \&
	\bibinfo{author}{Atat{\"u}re, M.}
	\newblock \bibinfo{title}{Frequency stabilization of the zero-phonon line of a
		quantum dot via phonon-assisted active feedback}.
	\newblock \emph{\bibinfo{journal}{Appl. Phys. Lett.}}
	\textbf{\bibinfo{volume}{105}}, \bibinfo{pages}{172107}
	(\bibinfo{year}{2014}).
	
	\bibitem{Bennett.2000}
	\bibinfo{author}{Bennett, C.~H.} \& \bibinfo{author}{DiVincenzo, D.~P.}
	\newblock \bibinfo{title}{Quantum information and computation}.
	\newblock \emph{\bibinfo{journal}{Nature}} \textbf{\bibinfo{volume}{404}},
	\bibinfo{pages}{247--255} (\bibinfo{year}{2000}).
	
	\bibitem{Kimble.2008}
	\bibinfo{author}{Kimble, H.~J.}
	\newblock \bibinfo{title}{The quantum internet}.
	\newblock \emph{\bibinfo{journal}{Nature}} \textbf{\bibinfo{volume}{453}},
	\bibinfo{pages}{1023--1030} (\bibinfo{year}{2008}).
	
	\bibitem{Ladd.2010}
	\bibinfo{author}{Ladd, T.~D.} \emph{et~al.}
	\newblock \bibinfo{title}{Quantum computers}.
	\newblock \emph{\bibinfo{journal}{Nature}} \textbf{\bibinfo{volume}{464}},
	\bibinfo{pages}{45--53} (\bibinfo{year}{2010}).
	
	\bibitem{Vamivakas.2009}
	\bibinfo{author}{Vamivakas, A.~N.}, \bibinfo{author}{Zhao, Y.},
	\bibinfo{author}{Lu, C.-Y.} \& \bibinfo{author}{Atat{\"u}re, M.}
	\newblock \bibinfo{title}{Spin-resolved quantum-dot resonance fluorescence}.
	\newblock \emph{\bibinfo{journal}{Nature Phys.}} \textbf{\bibinfo{volume}{5}},
	\bibinfo{pages}{198--202} (\bibinfo{year}{2009}).
	
	\bibitem{Ylmaz.2010}
	\bibinfo{author}{Y{\i}lmaz, S.~T.}, \bibinfo{author}{Fallahi, P.} \&
	\bibinfo{author}{Imamo{\u{g}}lu, A.}
	\newblock \bibinfo{title}{Quantum-dot-spin single-photon interface}.
	\newblock \emph{\bibinfo{journal}{Phys. Rev. Lett.}}
	\textbf{\bibinfo{volume}{105}} (\bibinfo{year}{2010}).
	
	\bibitem{Hogele.2012}
	\bibinfo{author}{H{\"o}gele, A.} \emph{et~al.}
	\newblock \bibinfo{title}{Dynamic nuclear spin polarization in the resonant
		laser excitation of an InGaAs quantum dot}.
	\newblock \emph{\bibinfo{journal}{Phys. Rev. Lett.}}
	\textbf{\bibinfo{volume}{108}}, \bibinfo{pages}{197403}
	(\bibinfo{year}{2012}).
	
	\bibitem{Chekhovich.2013}
	\bibinfo{author}{Chekhovich, E.~A.} \emph{et~al.}
	\newblock \bibinfo{title}{Nuclear spin effects in semiconductor quantum dots}.
	\newblock \emph{\bibinfo{journal}{Nature Mater.}}
	\textbf{\bibinfo{volume}{12}}, \bibinfo{pages}{494--504}
	(\bibinfo{year}{2013}).
	
	\bibitem{Yang.2013}
	\bibinfo{author}{Yang, W.} \& \bibinfo{author}{Sham, L.~J.}
	\newblock \bibinfo{title}{General theory of feedback control of a nuclear spin
		ensemble in quantum dots}.
	\newblock \emph{\bibinfo{journal}{Phys. Rev. B}} \textbf{\bibinfo{volume}{88}},
	\bibinfo{pages}{235304} (\bibinfo{year}{2013}).
	
	\bibitem{Luttjohann.2005}
	\bibinfo{author}{L{\"u}ttjohann, S.}, \bibinfo{author}{Meier, C.},
	\bibinfo{author}{Lorke, A.}, \bibinfo{author}{Reuter, D.} \&
	\bibinfo{author}{Wieck, A.~D.}
	\newblock \bibinfo{title}{Screening effects in InAs quantum-dot structures
		observed by photoluminescence and capacitance-voltage spectra}.
	\newblock \emph{\bibinfo{journal}{Appl. Phys. Lett.}}
	\textbf{\bibinfo{volume}{87}}, \bibinfo{pages}{163117}
	(\bibinfo{year}{2005}).
	
	\bibitem{Hosoda.1996}
	\bibinfo{author}{Hosoda, M.} \emph{et~al.}
	\newblock \bibinfo{title}{Rapid collapse of wannier-stark localization caused
		by space charge electric field screening in short-period superlattices}.
	\newblock \emph{\bibinfo{journal}{J. Appl. Phys.}}
	\textbf{\bibinfo{volume}{80}}, \bibinfo{pages}{5094} (\bibinfo{year}{1996}).
	
	\bibitem{Kroner.2008}
	\bibinfo{author}{Kroner, M.} \emph{et~al.}
	\newblock \bibinfo{title}{The nonlinear fano effect}.
	\newblock \emph{\bibinfo{journal}{Nature}} \textbf{\bibinfo{volume}{451}},
	\bibinfo{pages}{1022} (\bibinfo{year}{2008}).
	
	\bibitem{Hauck.2014}
	\bibinfo{author}{Hauck, M.} \emph{et~al.}
	\newblock \bibinfo{title}{Locating environmental charge impurities with
		confluent laser spectroscopy of multiple quantum dots}.
	\newblock \emph{\bibinfo{journal}{Phys. Rev. B}} \textbf{\bibinfo{volume}{90}},
	\bibinfo{pages}{235306} (\bibinfo{year}{2014}).
	
	\bibitem{Seidl.2005}
	\bibinfo{author}{Seidl, S.} \emph{et~al.}
	\newblock \bibinfo{title}{Absorption and photoluminescence spectroscopy on a
		single self-assembled charge-tunable quantum dot}.
	\newblock \emph{\bibinfo{journal}{Phys. Rev. B}} \textbf{\bibinfo{volume}{72}},
	\bibinfo{pages}{195339} (\bibinfo{year}{2005}).
	
	\bibitem{Warburton.2000}
	\bibinfo{author}{Warburton, R. J.} \emph{et~al.}
	\newblock \bibinfo{title}{Optical emission from a charge-tunable quantum ring}.
	\newblock \emph{\bibinfo{journal}{Nature}} \textbf{\bibinfo{volume}{405}},
	\bibinfo{pages}{926--929} (\bibinfo{year}{2000}).
	
	\bibitem{Strittmatter.2012}
	\bibinfo{author}{Strittmatter, A.} \emph{et~al.}
	\newblock \bibinfo{title}{Site-controlled quantum dot growth on buried oxide
		stressor layers}.
	\newblock \emph{\bibinfo{journal}{Phys. Stat. Sol.}}
	\textbf{\bibinfo{volume}{209}}, \bibinfo{pages}{2411--2420}
	(\bibinfo{year}{2012}).
	
	\bibitem{Strittmatter.2012b}
	\bibinfo{author}{Strittmatter, A.} \emph{et~al.}
	\newblock \bibinfo{title}{Lateral positioning of InGaAs quantum dots using a
		buried stressor}.
	\newblock \emph{\bibinfo{journal}{Appl. Phys. Lett.}}
	\textbf{\bibinfo{volume}{100}}, \bibinfo{pages}{48--51}
	(\bibinfo{year}{2012}).
	
	\bibitem{Warburton.2002}
	\bibinfo{author}{Warburton, R. J.} \emph{et~al.}
	\newblock \bibinfo{title}{Giant permanent dipole moments of excitons in
		semiconductor nanostructures}.
	\newblock \emph{\bibinfo{journal}{Phys. Rev. B}} \textbf{\bibinfo{volume}{65}},
	\bibinfo{pages}{113303} (\bibinfo{year}{2002}).
	
\end{thebibliography}
  
\end{document}